\documentclass[eqsecnum,showpacs,preprint,amsmath,aps,nofootinbib]{revtex4}  %
\usepackage{epsfig}
\usepackage{amssymb}
\usepackage[dvips]{color}
\usepackage[latin1]{inputenc}
\usepackage{graphicx}

\def\ud{\mbox{d}}

\def\beq{\begin{equation}}
\def\eeq{\end{equation}}
\def\beqq{\begin{eqnarray}}
\def\eeqq{\end{eqnarray}}

\newcommand{\bdm}{\begin{displaymath}}
\newcommand{\edm}{\end{displaymath}}

\def\pmb#1{\setbox0=\hbox{$#1$}%
  \kern-.025em\copy0\kern-\wd0
  \kern.05em\copy0\kern-\wd0
  \kern-.025em\raise.0433em\box0}

\makeatletter
\renewcommand*{\@fnsymbol}[1]{\ensuremath{\ifcase#1\or *\or \dagger\or
    \ddagger\or 
   \mathsection\or **\or \dagger\dagger
   \or \ddagger\ddagger \else\@ctrerr\fi}}
\makeatother

\begin{document}
\title{Perturbative evaluation of scalar two-point function
in the Cosmic Microwave Background power spectrum}

\author{Donato Bini}
\email[E-mail: ]{binid@icra.it}
\affiliation{Istituto per le Applicazioni del Calcolo
``M. Picone'', CNR, 00185 Rome, Italy\\
ICRA, University of Rome ``La Sapienza'',
00185 Rome, Italy\\
INFN, Sezione di Napoli, Complesso Universitario di Monte
S. Angelo, Via Cintia Edificio 6, 80126 Napoli, Italy}

\author{Giampiero Esposito}
\email[E-mail: ]{gesposit@na.infn.it}
\affiliation{Istituto Nazionale di Fisica Nucleare, Sezione di
Napoli, Complesso Universitario di Monte S. Angelo, 
Via Cintia Edificio 6, 80126 Napoli, Italy}

\date{\today}

\begin{abstract}
Recent work in the literature has found a suppression or, instead,
an enhancement of the Cosmic Microwave Background 
power spectrum in quantum gravity, although
the effect is too small to be observed, in both cases. The present
paper studies in detail the equations recently proposed for a
Born-Oppenheimer-type analysis of the problem. By using a perturbative
approach to the analysis of the nonlinear ordinary differential 
equation obeyed by the two-point function for scalar fluctuations, 
we find various explicit forms of such a two-point function, with the
associated power spectrum. In particular, a new family 
of power spectra is obtained and studied. The theoretical prediction
of power enhancement at large scales is hence confirmed.
\end{abstract}

\pacs{04.60.Ds}

\maketitle

\section{Introduction}

The attempts of building a quantum theory of gravity have given rise,
along the years, to substantial theoretical developments, e.g. the
discovery of ghost fields in the functional integral 
\cite{Feyn63, DeWi67, Fadd67}, Hawking radiation \cite{Hawk74, Hawk75}
and quantum field theory in curved spacetime \cite{DeWi75}, the
background-field method \cite{DeWi81, Abbott}, among the many. Quantum
gravity is currently expected to unify both the guiding principles
and all fundamental interactions of physics, although no agreement exists
on whether one should use field-theoretic or, instead, sharply different
structures (e.g. strings, branes, twistors, loops and spinfoams). Many
calculations in quantum gravity are very detailed and predictive, but
unfortunately the length scales and energies involved remain inaccessible
to experiments in the laboratories on earth, even though some encouraging
evidence exists that we might be approaching the era of quantum
gravity phenomenology \cite{Amelino}.

Over the last decades, however, the exciting (or puzzling) discoveries
in observational cosmology (e.g. dark matter, 
dark radiation, dark energy, the Cosmic Microwave Background 
(hereafter CMB) anisotropy spectrum) 
have led to several theoretical efforts, including 
the attempt of evaluating the effect of quantum gravity on the CMB power
spectrum. Interestingly, the work in Ref. \cite{Kief12} found that, in
canonical quantum gravity, a Jeffreys-Wentzel-Kramers-Brillouin analysis
of the Wheeler-DeWitt equation can yield a suppression of power at large
scales, in a model where a massive scalar field $\phi$ is coupled to a
spatially flat Friedmann-Lemaitre-Robertson-Walker universe, and
perturbations of $\phi$ are later considered. Another analysis of the
same set of nonlinear equations led in Ref. \cite{Bini13} to the opposite
prediction, i.e. an enhanced CMB power spectrum at large scales. 
Interestingly, a detailed application of 
Born-Oppenheimer methods \cite{Bertoni} to the same problem
has led, more recently, to calculations predicting again an enhanced 
power spectrum at large scales \cite{Kam13}. Since the
analysis in Ref. \cite{Kam13} avoids, by construction, all possible
inconsistencies related to unitarity violation \cite{Bertoni}, it has
been our aim to gain a deeper understanding of the potentialities of
the algorithm developed in Ref. \cite{Kam13}.

Section II summarizes recent results obtained from the 
Born-Oppenheimer technique \cite{BV} 
underlying the analysis in Ref. \cite{Kam13}. 
Section III studies the homogeneous equation 
associated to the nonlinear ordinary differential equation 
obeyed by the two-point function for scalar fluctuations.
The complete equation is studied in Sec. IV by means of a
perturbative ansatz. The resulting power spectrum is displayed
and discussed in Sec. V. Concluding remarks and open 
problems are presented in Sec. VI.
 
\section{Brief outline of recent results obtained from the 
Born-Oppenheimer method}
  
Following the work in Refs. \cite{Kief12,Bini13}, we study a
quantum cosmological model where a real-valued massive scalar
field is coupled to gravity in a spatially flat
Friedmann-Lemaitre-Robertson-Walker universe with scale factor $a$. 
Eventually, as shown in Ref. \cite{Kam13}, on considering the 
function $\rho$ which solves the Ermakov-Pinney equation 
\cite{Ermakov,Pinney,Lewis,Williams}
\begin{equation}
\left({{\rm d}^{2}\over {\rm d}\eta^{2}}+\omega^{2}\right)\rho
=\rho^{-3},
\end{equation}
where $\eta$ is the conformal-time variable such that 
$d\eta={dt \over a}$,
one arrives at building the normalized vacuum state (the prime
denoting derivative with respect to $\eta$) 
$$
(\pi \rho^{2})^{-{1\over 4}}
\exp \left[{{\rm i}\over 2} \int_{\;}^{\eta}
{{\rm d}{\tilde \eta}\over \rho^{2}}-{\nu^{2}\over 2}
\left({1\over \rho^{2}}-{\rm i}{\rho' \over \rho}\right)\right],
$$ 
and one derives a differential 
equation for the $2$-point function
$p(\eta)$ describing the spectrum of scalar fluctuations. Such an
equation reads as 
\begin{equation}
\label{eqtot}
\left[\frac{\ud^3 }{\ud \eta^3}+4\omega^2\frac{\ud }{\ud \eta}
+2\frac{\ud \omega^2}{\ud \eta}\right]p+\frac{F(\eta)}{m_P^2}=0,
\end{equation}
where $p(\eta)$ pertains to the vacuum state that reduces to the
Bunch-Davies vacuum \cite{Bunch} in the short 
wavelength regime, and $F$ is found to be
\begin{eqnarray}
F(\eta) & \equiv & -{{\rm d}^{3}\over {\rm d}\eta^{3}}
\left[{({p'}^{2}+4 \omega^{2}p^{2}-1)\over 4 {a'}^{2}}\right]
+{{\rm d}^{2}\over {\rm d}\eta^{2}}\left[{p'({p'}^{2}
+4\omega^{2}p^{2}+1)\over
4p {a'}^{2}}\right] \nonumber \\
&+& {{\rm d}\over {\rm d}\eta} \left \{{1\over 8 {a'}^{2}p^{2}}
\left[(1-4 \omega^{2}p^{2})^{2}+2{p'}^{2}(1+4 \omega^{2}p^{2})
+{p'}^{4}\right]\right \} \nonumber \\
&-& {\omega \omega'({p'}^{2}+4 \omega^{2}p^{2}-1)\over
{a'}^{2}}.
\end{eqnarray}

\section{The homogeneous differential equation}

The homogeneous differential equation associated with Eq. 
(\ref{eqtot}) is (see appendix A)
\begin{equation}
\label{eq_hom}
{\mathcal L}_{\omega}(p)\equiv \left(\frac{\ud^3 }{\ud \eta^3}
+4\omega^2\frac{\ud }{\ud \eta}
+2\frac{\ud \omega^2}{\ud \eta}\right)p=0.
\end{equation}
This third-order equation can be solved by writing $p$ in the form
\begin{equation}
p(\eta)=Y^{2}(\eta),
\end{equation}
and then considering the combination 
\begin{equation}
Z(\eta)=Y''+\omega^2 Y.
\end{equation}
Equation (\ref{eq_hom}) then becomes
\begin{equation}
6 Y' Z +2 Y Z'= 0\,,\qquad \Longrightarrow 
\qquad Z Y^3={\rm const}=C_{0},
\end{equation}
so that one gets the $1$-parameter family of second-order ordinary 
differential equations
\begin{equation}
\label{fundam}
\boxed{
Y^3 (Y''+\omega^2 Y)=C_{0}.
}
\end{equation}
In fact pointing out that 
\begin{equation}
p'=2YY', \qquad p''=2Y'^{2}+2YY'', \qquad
p'''=2YY'''+6Y'Y'' ,
\end{equation}
the third-order equation (\ref{eq_hom}) is re-expressed as
\begin{equation}
{\mathcal L}_\omega (Y^2)=
2 \Bigr(YY'''+3Y'Y''+4\omega^{2}YY'
+2\omega \omega'Y^{2}\Bigr)=0 \,.
\end{equation}
Multiplying both sides of this equation by $Y^2$ we obtain the 
equivalent form
\begin{equation}
0=Y^2 {\mathcal L}_\omega (Y^2)=
2[Y^3 Y''+\omega^2 Y^4]'=2[Y^3 Z ]' .
\end{equation}
In other words, $Y^2 {\mathcal L}_\omega (Y^2)$ vanishes and is itself 
proportional to the derivative of  $Y^3 (Y''+\omega^2 Y)$, so that
Eq. (\ref{fundam}) follows easily.

Therefore, as soon as a special choice of $\omega(\eta)$ is made, 
Eq. (\ref{fundam}) can be  solved for $Y(\eta)$.
For example, when the Hubble parameter $H$ is constant, a natural choice 
for $\omega^2$ is the following \cite{Kam13}:
\begin{equation}
\omega^2=k^2\left(1-\frac{2}{k^2\eta^2}\right),
\end{equation}
which implies that $\omega \to k$ 
as soon as the conformal time goes to infinity.
It is then convenient to introduce the new variable 
$x \equiv -k\eta$ and rescale 
the constant $C_0=k^2 B_0$.
One finds for $Y(x)$ the following solution (depending on three arbitrary 
integration constants):
\begin{equation}
\label{fin_sol}
Y^2=\frac{B_0}{C_1}Y_-^2+C_1(2C_2 Y_-+Y_+)^{2},
\end{equation}
where
\begin{equation}
Y_-(x)=\cos x -\frac{\sin x}x=-\sqrt{\frac{\pi x}{2}}J_{3/2}(x)  \,,\quad
Y_+(x)=\sin x  +\frac{\cos x}x =\sqrt{\frac{\pi x}{2}}J_{-3/2}(x) ,
\end{equation}
and only $Y_{-}(x)$ has a finite limit at $x=0$. 
After a suitable redefinition of constants, one finds then for $Y^2$ the 
three elementary solutions
\begin{equation}
\label{gen_y_quad0}
Y^2=c_1 Y_{+}^{2} +c_2 Y_{-}^{2} +c_3 Y_{+} Y_{-} .
\end{equation}
This class of solutions is rich enough and contains either combination of 
Bessel-$J$ functions and polynomials.
In fact, when $c_3=2\sqrt{c_1c_2}$, Eq. (\ref{gen_y_quad0}) reduces to
\begin{equation}
\label{gen_y_quad}
Y^2=(\sqrt{c_1} Y_+ +\sqrt{c_2} Y_-)^{2}.
\end{equation}
Similarly, when $c_3=0$ and $c_1=c_2=1$ we have
\begin{equation}
Y^2=Y_{-}^{2}+Y_{+}^{2}=1+\frac{1}{x^2}.
\end{equation}
In particular, following Ref. \cite{Kam13}, 
the choice of constants  $c_1=1/2$, $c_2=0=c_3$, which implies
\begin{equation}
Y(x)=\frac1{\sqrt{2}} Y_{+}(x),
\end{equation}
should be preferred.

In view of its simplicity and of its relevance for a
pure de Sitter expansion \cite{Kam13}, 
in the following perturbative analysis, we will 
use as solutions of the homogeneous equation (\ref{eq_hom})
\begin{equation}
\label{special_sol}
p(x)=\frac1{2k} \left(1+\frac{1}{x^2}\right)\equiv p_{0}(x) 
\,,\qquad \omega(x)
=k \sqrt{1-\frac{2}{x^2}}\equiv \omega_{0}(x),
\end{equation}
where we have restored for convenience the original variables 
$p$ and $\omega$. In terms of the auxiliary variable defined in 
Eq. (3.2), such a solution satisfies Eq. (3.5) with 
$C_{0}={1\over 4}$ which is associated, in turn, with the chosen
initial data for $Y(x)$ and $\omega(x)$ at $x=0$; in fact
$C_{0}=Y^{3}(0)[Y''(0)+\omega^{2}(0)Y(0)]$. Our choice of initial
conditions results from the request (as in Ref. \cite{Kam13},
Eqs. (32) and (33) therein) to reproduce the de Sitter result
in absence of quantum corrections.

\section{The complete equation}

Since it is rather difficult to solve exactly the complete 
equation (2.5), we now look for some specific 
conditions that make it possible to find a perturbative solution. 
In a viable single-field inflationary model, one has an evolution of 
cosmological perturbations based on the slow-roll paradigm. 
However, in order to illustrate the main effect of quantum gravity on the 
spectrum, it is sufficient to neglect slow-roll parameters, which
leads to a pure de Sitter expansion $a(t)=e^{Ht}$ with constant $H$,
for which the condition ${da\over dt}=Ha$, re-expressed through 
conformal time $\eta$, becomes
\begin{equation}
{1\over a}{{\rm d}a \over {\rm d}\eta}=Ha \Longrightarrow 
{\rm d} \left({1\over a}\right)=-H \; {\rm d} \eta \Longrightarrow
a=-{1\over H \eta}.
\end{equation}
Besides $x \equiv -k\eta$ we also introduce the rescaled variables
\begin{equation} 
\Omega \equiv \frac{\omega}{k},
\; \quad P \equiv kp,
\end{equation}
as well as the (small) quantity 
\begin{equation}
\varepsilon \equiv \frac{H^2}{m_{P}^{2} k^{3}}.
\end{equation}

We undertake now the analysis of solutions which perturb the special one 
given in Eq. (\ref{special_sol}), that we write in the form
\begin{equation}
\label{pert_sol2}
P(x)=\frac12 \left(1+\frac1{x^{2}}\right)+\varepsilon P_1 (x), \; \qquad  
\Omega^2(x)= 1-\frac2{x^{2}} +\varepsilon W_1 (x),
\end{equation}
where the unperturbed solutions $P_0$ and $\Omega_0$  
have been introduced before, in (\ref{special_sol}).
Note that, by using for $P_0$ either $Y_{+}^{2}$, 
$Y_{-}^{2}$ or $Y_{+}Y_{-}$ (or a linear combination of them)
instead of the simple solution (4.4) results only in some 
mathematical complications. For example, one finds the 
cumbersome equation (B1) in appendix B.
By using for $P_0(x)$ (i.e., for $Y_0(x)$) and $\Omega_0(x)$ 
the expressions given in (4.4), 
Eq. (B1) is much simplifid and reduces to
\begin{equation}
-P_1'''-4\left(1-\frac{2}{x^2}\right)P_1'-\left(1+\frac{1}{x^2}\right)
W_1'  -\frac{8}{x^3}P_1 +\frac{4}{x^3}(W_1 -1)=0.
\end{equation}
A simple inspection of this equation suggests introducing
\begin{equation}
{\widetilde W}_{1} \equiv W_{1}-1,
\end{equation}
so that the final equation for perturbative quantities is given by
\begin{equation}
-x^3 P_1'''-4x\left(x^2-2\right)P_1'-8 P_1
-x\left(x^2+1\right){\widetilde W}_1'   
+4 {\widetilde W}_1 =0,
\label{(4.7)}
\end{equation}
that is, recalling the definition of the operator ${\mathcal L}_{\Omega}$ 
of Eq. (\ref{eq_hom}) 
\begin{equation}
{\mathcal L}_{\Omega_0}(P_1)  =
-\left(1+\frac{1}{x^2}\right){\widetilde W}_1'   
+ \frac{4}{x^3} {\widetilde W}_1\equiv Q(x) ,
\end{equation}
$Q(x)$ being our notation for the inhomogeneous term, so that
\begin{equation}
{\widetilde W}_{1}(x) =\left( \frac{x^2}{1+x^2}\right)^{2} 
\left(C_1-\int^x \frac{Q(z)(z^2+1)}{z^2} {\rm d}z \right).
\end{equation}
This equation can be solved formally and the result is as follows: 
\begin{eqnarray}
\label{sol_formal_G}
P_1(x)
&=& \int^x  Q(z)G(x,z) {\rm d}z+c_{1} Y_{-}^{2}+c_{2}Y_{+}^{2}
+c_{3} Y_{+} Y_{-} ,
\end{eqnarray}
where the Green function is given by
\begin{equation}
G(x,z)=\frac{1}{2}\left(Y_{+}(z) Y_{-}(x)-Y_{+}(x) 
Y_{-}(z)\right)^{2} ,
\end{equation}
and the last three terms represent the general solution of the associated 
homogeneous equation. Note that 
\begin{equation}
\label{identities_for_G}
G(x,x)=0\,,\quad \partial_x G(x,z)|_{z=x}=0\,,\quad 
\partial_{xx} G(x,z)|_{z=x}=1.
\end{equation}
Equation (\ref{sol_formal_G}) can be proven as 
follows (omitting the solution 
of the homogeneous equation for which it is simply 
${\mathcal L}_{\Omega_0}(c_1 Y_-^2+c_2Y_+^2+c_3 Y_+Y_-)=0$).
Using Eqs. (\ref{identities_for_G}) we have immediately
\begin{eqnarray}
\frac{{\rm d}}{{\rm d}x}P_1 &=& 
Q(x)G(x,x)+ \int^x  Q(z)\partial_x G(x,z) {\rm d}z
= \int^x  Q(z)\partial_x G(x,z) {\rm d}z,
\nonumber\\
\frac{{\rm d}^2}{{\rm d}x^2}P_1 &=&  
Q(x)\partial_x G(x,z)|_{z=x}
+\int^x  Q(z)\partial_{xx} G(x,z) {\rm d}z
=\int^x  Q(z)\partial_{xx} G(x,z) {\rm d}z,
\nonumber\\
\frac{{\rm d}^3}{{\rm d}x^3}P_1 &=&  
Q(x)\partial_{xx} G(x,z)|_{z=x}
+\int^x  Q(z)\partial_{xxx} G(x,z) {\rm d}z 
\nonumber \\
&=& Q(x)+\int^x  Q(z)\partial_{xxx} G(x,z) {\rm d}z,
\end{eqnarray}
hence the sought for result
\begin{equation}
{\mathcal L}_{\Omega_0}(P_1)= Q(x)+ \int^x  Q(z) 
[{\mathcal L}_{\Omega_0} G(x,z)] {\rm d}z=Q(x).
\end{equation}
In spite of this nice result for the general representation of $P_1$, 
however, we are going to consider special situations in which 
$Q(x)$ is given. 

For example, we can find a series solution consistently for $P_1$ 
and ${\widetilde W}_1$.  
It is worth discussing separately the following simple cases.
\begin{itemize}
  \item Case $P_1=0$.
In this case we have
\begin{equation}
-x\left(x^2+1\right){\widetilde W}_1'   
+4 {\widetilde W}_1 =0,
\end{equation}
with solution
\begin{equation}
{\widetilde W}_1 = C_1\left( \frac{x^2}{1+x^2}\right)^{2}.
\end{equation}
  \item Case ${\widetilde W}_1=0$.
In this case we have the homogeneous equation 
\begin{equation}
x^3 P_1'''+4x\left(x^2-2\right)P_1'+8 P_1=0,
\end{equation}
with the known solution
\begin{eqnarray}
P_1 &=& c_1 Y_-^2+c_2Y_+^2+c_3 Y_+Y_-.
\end{eqnarray}
\item Series solution for both $P_1$ and ${\widetilde W}_1$.

Looking for solutions having the form
\begin{equation}
P_1=\sum_{k=0}^{n_1} A_k x^k\,,\qquad {\widetilde W}_1
=\sum_{k=0}^{n_2} B_k x^{k},
\end{equation}
a particular solution involving a minimum number of coefficients 
$A_k$ and $B_k$ (\lq\lq minimal solution") is given by
\begin{equation}
\label{eq:4.21}
P_1= A_0+A_2x^2\,,\qquad  {\widetilde W}_1= 2A_0-4A_2x^{2},
\end{equation}
with $n_1=n_2=2$ and $A_0$ and $A_2$ undetermined constants. 
Note that, even though in these solutions there appear powers of
the time variable $x$, which imply a more rapid growth of the
perturbation itself, there is enough room for the study presented
here. In fact, the condition for obtaining perturbations that can
be thrusted can be expressed by the majorization
$$
\left | \varepsilon {P_{1} \over {\widetilde W}_{1}} \right | << 1,
$$
and the very small values of $\varepsilon$ (getting smaller at higher
wave numbers $k$) allow anyway for polynomial variations of the
time variable, even of degree much larger than $2$.

\end{itemize}

\section{The power spectrum}

For the $3$ cases considered in Sec. IV we can now evaluate 
the power spectrum ${\cal P}_{\nu}$, given by \cite{Kam13}
\begin{equation}
{\cal P}_{\nu}={k^{3}\over 2\pi^{2}}p=\left(\frac{k}{2\pi}  
\right)^2 2 P \equiv {\cal P}_* \, 2P.
\label{(5.1)}
\end{equation}
For example, in the $3$ perturbative cases considered above we have
\begin{itemize}
\item Case $P_1=0$
\begin{equation}
\label{P1_eq_0}
{\cal P}_{\nu}={\cal P}_* \left(1+\frac{1}{x^2}  \right).
\end{equation}

\item Case ${\widetilde W}_1=0$
\begin{equation}
\label{W1_eq_0}
{\cal P}_{\nu}={\cal P}_* \left\{
1+\frac{1}{x^2} +2 \varepsilon \left[
c_1Y_-^2+c_2Y_+^2+c_3 Y_+Y_-\right] 
\right\}.
\end{equation}
The additional term of first order in $\varepsilon$
has the following MacLaurin expansion: 
\begin{equation}
c_2\left(1+\frac{1}{x^2}\right)-\frac{c_3}{3}
\left( 1-\frac{2}{5}x^2\right)x+{\rm O}(x^4).
\end{equation}
\item  Series solution for both $P_1$ and ${\widetilde W}_1$.
Last, but not least, the opportunities offered by Eq. (\ref{eq:4.21}),
are richer because, even in the case of the  \lq\lq minimal solution",   
the resulting power spectrum depends on a pair of arbitrary 
constants and reads as
\begin{equation}
\label{caso_part1}
{\cal P}_{\nu}={\cal P}_*\left[1+\frac{1}{k^{2}\eta^{2}}
+ 2\varepsilon  
(A_{0}+A_{2}k^{2}\eta^{2})\right].
\end{equation}
Here, the term proportional to $A_{0}$ scales as $k^{-3}$ and
leads to an increase of power for large scales, as in 
Ref. \cite{Kam13}.

\end{itemize}

In all previous cases the dependence of the power spectrum on 
several constants can be used to fit experimental data. Such data, 
however, are far beyond the actual sensitivity of existing devices.
More precisely, in Ref. \cite{Bini13} it has been shown that
the quantum-gravitationally corrected Schr\"odinger equation leads to a
modification (to first order in $\varepsilon$) of the power spectrum 
by a correction function $C_k$,
such that one can translate this modification also to the standard
power spectrum in the following way: 
\begin{equation}
\label{pow_pert}
{\cal P}_{\nu}^{(1)}(k)={\cal P}_{\nu}^{(0)}(k)\,C_{k}^{2}\,.
\end{equation}
One can write
\begin{equation} \label{Ckdelta}
C_{k}^{2}=1+\delta^\pm_{\mathrm{WDW}}(k) + {\rm O}(\varepsilon^2) , 
\end{equation}
where $\delta^\pm_{\mathrm{WDW}}(k)$ either takes the form 
\begin{equation} \label{delta+}
\delta^+_{\mathrm{WDW}}(k)=179.09 \varepsilon ,
\end{equation}
or the form 
\begin{equation} \label{delta-}
\delta^-_{\mathrm{WDW}}(k)=-247.68 \varepsilon . 
\end{equation}
Note that we can also cast our result in a form similar to that 
given by Eq. (\ref{pow_pert}). In fact, for a general power spectrum 
of the form
\begin{equation}
{\cal P}_{\nu} = {\cal P}_{\nu}^{(0)}(k)
\left(1+ \varepsilon \frac{ k^{2} \eta^{2}}{(1+k^{2} \eta^{2})} 
{\mathcal F}_{(k,\eta)} \right)+ {\rm O}(\varepsilon^2),
\end{equation}
from which the identification
\begin{equation}
C_{k}^{2}=1 +\delta(k,\eta)+ {\rm O}(\varepsilon^2),
\end{equation}
where
\begin{equation}
\delta(k,\eta) \equiv \varepsilon \frac{ k^2\eta^2  }
{(1+k^{2} \eta^{2})} {\mathcal F}_{(k,\eta)}.
\end{equation}
For instance, in the case (\ref{caso_part1}) discussed above we have
\begin{equation}
{\mathcal F}_{(k,\eta)}= 2 (A_0+A_2 k^2 \eta^2),
\end{equation}
and then
\begin{equation}
\delta(k,\eta)= \varepsilon \frac{2 k^2\eta^2  }
{(1+k^{2} \eta^{2})} (A_0+A_2 k^2 \eta^2).
\end{equation}
In order to compare $\delta(k,\eta)$ with its WDW counterparts,  
say $\delta_{\mathrm{WDW}}(k)=C \varepsilon$ where $C$ is a constant, 
one can compute for example $\delta(k,\eta)$ at a certain value 
$k_* \eta_*$ properly chosen (e.g., the value $\eta_*$ which extremizes 
$\delta(k,\eta)$); this choice, as well as other similar choices, leads 
to a  $\delta(k_*,\eta_*)$ which still depends upon $A_0$ and $A_2$ and 
such parameters can be adjusted to fit experimental data, for example the 
(slow-roll) parameters of any inflationary model. It is worth
stressing that the factor multiplying $\varepsilon$ on the 
right-hand side of (5.14) plays the role of modulating factor.
This concept will be discussed again below.

The basic equations in the theory of the spectral index $n_s$ 
and its running $\alpha_s$ involve the slow-roll parameters
\cite{calcagni} 
$
\eta \equiv -{\ddot \phi \over H {\dot \phi}}, \;
\epsilon \equiv -{{\dot H}\over H^{2}}
=2 {{\dot \phi}^{2}\over H^{2}}, \;
\Xi^2 \equiv
{1\over H^{2}}{{\rm d}\over {\rm d}t}{\ddot \phi \over {\dot \phi}},
$
where we have used $8 \pi G=1$ units in the formula
for $\epsilon$. Therefore 
\begin{equation}
n_{s}-1 \equiv {{\rm d}\log {\cal P}_{\nu} \over 
{\rm d}\log k} \approx 2\eta -4\epsilon -3 \delta^\pm_{\mathrm{WDW}}
\end{equation}
and
\begin{equation}
\alpha_{s} \equiv {{\rm d}n_{s} \over {\rm d}\log k}
\approx 2 (5 \epsilon \eta-4 \epsilon^{2}-\Xi^{2})
+9 \delta^\pm_{\mathrm{WDW}}\,,
\end{equation}
where use has been made of the approximate formula 
\begin{equation}
{{\rm d}\over {\rm d}\log k} \approx {1\over H}
{{\rm d}\over {\rm d}t}\,,
\end{equation}
jointly with the equations of motion.

The work in Ref. \cite{Bini13} has shown that the absolute value
of $\delta_{{\rm WDW}}^{\pm}$ is majorized by numbers of order
$10^{-12}$ when $k$ is replaced by $k/k_{{\rm min}}$, where 
$k_{{\rm min}}$ is the largest observable scale, or by numbers
of order $10^{-9}$ when $k$ is replaced by $k/k_{0}$, where
$k_{0}$ is the pivot scale used in the WMAP9 analysis. 
By comparing such quantum-gravitational corrections to the spectral
index $n_s$ and its running $\alpha_s$ derived above with the values
determined from the WMAP9 data, $n_s = 0.9608 \pm 0.0080$ and $\alpha_s
= -0.023 \pm 0.011$ (using the WMAP9+eCMB+BAO+$H_0$ dataset in 
both cases) \cite{Hinshaw12},  and the 2013 results of the Planck mission, 
$n_s = 0.9603 \pm 0.0073$ and $\alpha_s
= -0.013 \pm 0.009$ (using additionally the WMAP polarization data in both 
cases) \cite{Planck13}, one sees that the corrections in \cite{Bini13}
are completely drowned out by the statistical uncertainty
in the data. We are currently trying to understand whether one can
arrive at formulas where $\delta_{{\rm WDW}}$ is systematically
replaced by our $\delta(k,\eta)$. 

\subsection{Special solutions of the perturbative equation}

It is easy to find special solutions to the final perturbative equation
with more terms. For example,
looking for a sixth-order polynomial solution of Eq. (\ref{(4.7)})
\begin{eqnarray}
\label{eq:4.21bis}
P_{1}^{(6)}&=&\frac12 B_0 +\left(\frac54 B_5-\frac38 B_3\right)x-\frac14
B_2x^2+\left(-\frac14 B_3+\frac12 B_5\right)x^{3} \nonumber \\
&+& \left(-\frac14 B_4+\frac98
B_6\right)x^4-\frac14 B_5x^5-\frac14 B_6x^{6},
\end{eqnarray}
\begin{equation}
{\widetilde W}_{1}^{(6)}=B_0+B_2x^2+B_3x^3+B_4x^4+B_5x^5+B_6x^6,
\end{equation}
where the $B_k$ are arbitrary constants, we have 
\begin{equation}
\label{pert_sol2}
P(x)=\frac12 \left(1+\frac1{x^{2}}\right)
+\varepsilon P_{1}^{(6)} (x), \; \qquad
\Omega^2(x)= 1-\frac2{x^{2}} +\varepsilon 
(1+\widetilde W_{1}^{(6)} (x)),
\end{equation}
and the resulting power spectrum is given by
\begin{equation}
\frac{{\cal P}_{\nu}}{ {\cal P}_*}=1+\frac1{x^{2}} 
+2\varepsilon  P_{1}^{(6)} (x).
\end{equation}
Another (perhaps more interesting) solution should involve also
negative powers of $x$. For this purpose, we look for solutions
in the form
\begin{equation}
P_{1}^{(-3,3)}=\sum_{k=-3}^{3}A_{k}x^{k}, \;
{\widetilde W}_{1}^{(-3,3)}=\sum_{k=-3}^{3}B_{k}x^{k},
\label{(5.23)} 
\end{equation}
where there is full freedom to choose the lower and upper summation
limits. For simplicity, we have here set the limits 
equal to $-3$ and $3$,
respectively. By inserting such an ansatz into Eq. (4.7) we find
\begin{eqnarray}
P_{1}^{(-3,3)}&=& -{1\over 4}{B_{-3}\over x^{3}}
-{1\over 2}{B_{0}\over x^{2}}-{1\over 4}\left(B_{1}
+{3\over 2}B_{3}\right)x-{B_{2}\over 4}x^{2}
-{B_{3}\over 4}x^{3},
\nonumber \\
& \; & {\widetilde W}_{1}^{(-3,3)}={B_{-3}\over x^{3}}
-{B_{1}\over x}+B_{0}+B_{1}x+B_{2}x^{2}+B_{3}x^{3},
\end{eqnarray}
where compatibility requires that $B_{-1}=-B_{1}$ and
$A_{0}$ can be always set to $0$. An interesting particular case of
such a framework corresponds to choosing $B_{1}=B_{2}=B_{3}=0$.
Our previous formula reduces then to
\begin{eqnarray}
P_{1}^{(-3,3)}&=& -{1\over 4}{B_{-3}\over x^{3}}
-{1\over 2}{B_{0}\over x^{2}},
\nonumber \\
& \; & {\widetilde W}_{1}^{(-3,3)}={B_{-3}\over x^{3}}
-{B_{1}\over x}+B_{0}.
\end{eqnarray}
Note that the result (33) in Ref. \cite{Kam13} is described by the
simple choice $B_{-3}=0,B_{0}=-2$.

Our power spectrum can be written in the form 
\begin{equation}
\frac{{\cal P}_{\nu}}{ {\cal P}_*}=1+\frac1{x^{2}} 
-{\varepsilon \over x^{2}}\left({B_{-3}\over x}+B_{0}\right).
\end{equation}
This implies, with the notation used in (5.7), that
\begin{equation}
\delta_{{\rm WDW}}=-\varepsilon {\Bigr(B_{-3}+B_{0}x \Bigr)\over
x(x^{2}+1)}.
\end{equation}
As we have said earlier, it is interesting to look at the structure
of the modulating factor in the formula expressing the enhancement
of the power spectrum, i.e., the ratio
${\delta_{{\rm WDW}}\over \varepsilon}$ in the above formula. 
A nice feature of such a ratio is its summability on the whole real
line, provided one adopts the principal-value prescription for the integral
including the origin.
With this understanding, one can evaluate its average, which is
then equal to $\pi B_{0}$. This makes it possible to compare straight
away our $\langle \delta_{{\rm WDW}} \rangle$ with other values,
whether or not existing in the literature.  

\section{Concluding remarks and open problems}

In our paper we have applied a perturbative technique
for the evaluation of the scalar two-point function in the CMB power
spectrum, relying upon the general technique developed in Ref. 
\cite{Kam13} for de Sitter evolution,
with the associated fundamental equations (2.2) and
(2.3). Our results in Sec. IV, elegant and at the same time simple in 
their derivation, are entirely original and lead to the
theoretical formulas for the power spectrum displayed in Sec. V which 
may depend on several parameters.
Such formulas reduce to the existing ones \cite{Kam13} in a 
particular case, but have better potentialities because the lower and
upper limit of summation in (\ref{(5.23)}) are arbitrary. Hence one might
arrive at more accurate theoretical predictions, to be hopefully checked 
against observations. 

Recently, the work in Ref. \cite{Kam14}, relying upon Ref. \cite{Kam13},
has evaluated the spectra of scalar and tensor perturbations to
first order in the slow-roll approximation, which has been found to
provide qualitatively new quantum gravitational effects with respect
to the pure de Sitter case. We think it would also be interesting to 
apply our technique to the slow-roll phase studied therein.

\acknowledgments 
The authors are indebted to Claus Kiefer and Manuel Kr\"{a}mer
for enlightening correspondence, and to F. Pessina for useful 
discussions at the beginning of the present project.
G. E. is grateful to the Dipartimento di Fisica of
Federico II University, Naples, for hospitality and support. 

\appendix

\section{The operator ${\mathcal L}_{\omega}$}

The linear differential operator defined in Eq. (3.1) is not a
derivation and hence does not obey the Leibniz rule, 
but satisfies instead the following property:
\begin{equation}
{\mathcal L}_{\omega}(fg)=f{\mathcal L}_{\omega}(g)
+g{\mathcal L}_{\omega}(f)+3(f'g')'-2(\omega^2)' fg.
\end{equation}
In the case $f=g$ this relation implies
\begin{equation}
{\mathcal L}_{\omega}(f^2)=2f{\mathcal L}_{\omega}(f)
+3(f'{}^2)'-4\omega \omega' f^{2} .
\end{equation}
Another useful relation which follows from those considered above  
states that
\begin{equation}
{\mathcal L}_{\omega}((f-g)^2)=2(f-g) [{\mathcal L}_{\omega}(f)
-{\mathcal L}_{\omega}(g)]+6(f'-g')(f''-g'')-4\omega \omega' (f-g)^{2}.
\end{equation}

\section{Perturbations of solutions of the homogeneous
equation; the composite function $F(\eta(x))$}

The cumbersome equation mentioned after Eq. (4.4) reads,
explicitly, as
\begin{eqnarray}
\; & \; & {{\rm d}^{3}\over {\rm d}x^{3}}P_{1}(x)=-4P_{1}(x)
\Omega_{0}(x){{\rm d}\over {\rm d}x}\Omega_{0}(x)
\nonumber \\
&+& \left[-4 Y_{0}^{2}(x)\left({{\rm d}\over {\rm d}x}
\Omega_{0}(x)\right)-16 \Omega_{0}(x)Y_{0}(x)
\left({{\rm d}\over {\rm d}x}Y_{0}(x)\right)\right]\Omega_{1}(x)
\nonumber \\
&-& 4 \Omega_{0}^{2}(x)\left({{\rm d}\over {\rm d}x}P_{1}(x)\right)
+\left({x^{4}\over Y_{0}^{3}(x)}(1-4C_{0})
+48 x^{2} Y_{0}(x)\right)
\left({{\rm d}\over {\rm d}x}Y_{0}(x)\right)^{3}
\nonumber \\
&+& \left[-16 x^{3} Y_{0}^{2}(x)\Omega_{0}^{2}(x)
+20 x^{4}Y_{0}^{2}(x)\Omega_{0}(x)
\left({{\rm d}\over {\rm d}x}\Omega_{0}(x)\right)
+24 x Y_{0}^{2}(x)\right]
\left({{\rm d}\over {\rm d}x}Y_{0}(x)\right)^{2}
\nonumber \\
&+& \left[112 x^{3}Y_{0}^{3}(x)\Omega_{0}(x)
\left({{\rm d}\over {\rm d}x}\Omega_{0}(x)\right)
+{x^{4}\over Y_{0}^{5}(x)}
\left({1\over 2}-12C_{0}^{2}\right) \right.
\nonumber \\
&+& {x^{4}\Omega_{0}^{2}(x)\over Y_{0}(x)}
+16 x^{4} Y_{0}^{3}(x)\Omega_{0}(x)
\left({{\rm d}^{2}\over {\rm d}x^{2}}\Omega_{0}(x)\right)
-4{x^{4}\Omega_{0}^{2}(x)C_{0}\over Y_{0}(x)}
+72{x^{2}C_{0}\over Y_{0}(x)}
\nonumber \\
&-& \left . {6x^{2}\over Y_{0}(x)}
+48 x^{2}Y_{0}^{3}(x)\Omega_{0}^{2}(x)
+16 x^{4}Y_{0}^{3}(x)
\left({{\rm d}\over {\rm d}x}\Omega_{0}(x)\right)^{2}
+{x^{4}C_{0}\over Y_{0}^{5}(x)}\right]
\left({{\rm d}\over {\rm d}x}Y_{0}(x)\right)
\nonumber \\ 
&-& 4 Y_{0}^{2}(x)\Omega_{0}(x)
\left({{\rm d}\over {\rm d}x}\Omega_{1}(x)\right)
+24{x^{3}C_{0}^{2}\over Y_{0}^{4}(x)}
-4{x^{3}C_{0}\over Y_{0}^{4}(x)} 
\nonumber \\
&-& 16 x^{3}C_{0}\Omega_{0}^{2}(x)-6x +24 x Y_{0}^{4}(x)
\Omega_{0}^{2}(x) 
\nonumber \\
&+& 6 x^{4}Y_{0}^{4}(x)\left({{\rm d}\over {\rm d}x}\Omega_{0}(x)
\right) \left({{\rm d}^{2}\over {\rm d}x^{2}}\Omega_{0}(x)
\right)+8x^{3}\Omega_{0}^{2}(x)
-{1\over 2}{x^{3}\over Y_{0}^{4}(x)}
\nonumber \\
&-& 12 x^{4}Y_{0}^{4}(x)\Omega_{0}^{3}(x)
\left({{\rm d}\over {\rm d}x}\Omega_{0}(x)\right)
-16 x^{3}Y_{0}^{4}(x)\Omega_{0}^{4}(x) 
\nonumber \\
&+& 24 x^{3}Y_{0}^{4}(x)\Omega_{0}(x)
\left({{\rm d}^{2}\over {\rm d}x^{2}}\Omega_{0}(x)\right)
+72 x^{2}Y_{0}^{4}(x)\Omega_{0}(x)
\left({{\rm d}\over {\rm d}x}\Omega_{0}(x)\right)
\nonumber \\
&+& 24 x^{3}Y_{0}^{4}(x)
\left({{\rm d}\over {\rm d}x}\Omega_{0}(x)\right)^{2}
+2 x^{4}Y_{0}^{4}(x)\Omega_{0}(x)
\left({{\rm d}^{3}\over {\rm d}x^{3}}\Omega_{0}(x)\right)
\nonumber \\
&+& 2 x^{4}\Omega_{0}(x)\left({{\rm d}\over {\rm d}x}
\Omega_{0}(x)\right)+4C_{0}x^{4}\Omega_{0}(x)
\left({{\rm d}\over {\rm d}x}\Omega_{0}(x)\right),
\end{eqnarray}
where $P_{0}(x)=Y_{0}^{2}(x)$ and $\Omega_{0}(x)$ 
are generic solutions of the homogeneous
equation as studied in Sec. III.

If $P_{1}$ vanishes, the composite function $F(\eta(x))$ is
obtained from the general formulae (2.3), (4.1)--(4.4) 
through the formula
\begin{equation}
F(\eta(x))[P_{1}=0]=\varepsilon m_{P}^{2}k^{2}
{\widetilde F}(\eta(x))[P_{1}=0]
+{\rm O}(\varepsilon^{2}),
\end{equation}
having set
\begin{eqnarray}
{\widetilde F}(\eta(x))[P_{1}=0]& \equiv & 
\left \{ {6\over x^{5}}
+{1\over 2}{{\rm d}^{2}\over {\rm d}x^{2}}\Bigr[(1+x^{2})^{-1}(-x^{-3}
-3x^{-1}+2x^{3})\Bigr] \right .
\nonumber \\
&-& {1\over 2}{{\rm d}\over {\rm d}x}\Bigr[(1+x^{2})^{-2}
(x^{-4}(2+3x^{2})^{2}+2x^{2}(2-3x^{-4}-2x^{-6})+x^{-4})\Bigr]
\nonumber \\
&-& \left . {2(1+3x^{2})\over x^{5}} \right \},
\end{eqnarray}
where
\begin{equation}
{{\rm d}^{2}\over {\rm d}x^{2}}\Bigr[(1+x^{2})^{-1}
(-x^{-3}-3x^{-1}+2x^{3})\Bigr]
=(1+x^{2})^{-3}\Bigr[-12x^{-5}-40x^{-3}-48x^{-1}-24x-4x^{3}\Bigr],
\end{equation}
\begin{eqnarray}
\; & \; & 
-{{\rm d}\over {\rm d}x}\Bigr[(1+x^{2})^{-2}(x^{-4}(2+3x^{2})^{2}
+2x^{2}(2-3x^{-4}-2x^{-6})+x^{-4})\Bigr] 
\nonumber \\
&=& (1+x^{2})^{-3}\Bigr[4x^{-5}+20x^{-3}+36x^{-1}+28x+8x^{3}\Bigr],
\end{eqnarray}
which lead to
\begin{equation}
F(\eta(x))[P_{1}=0]=-{4 \varepsilon m_{P}^{2}k^{2} \over x^{3}}
+{\rm O}(\varepsilon^{2}).
\end{equation}
Interestingly, we also find the simple but nontrivial equality
\begin{equation}
{\widetilde F}(\eta(x))[P_{1}=0]
={\widetilde F}(\eta(x))[W_{1}=0]
={\widetilde F}(\eta(x))[P_{1} \not=0,W_{1} \not=0],
\end{equation}
which means that, to first order in $\varepsilon$, 
the value taken by $F(\eta(x))$ is unaffected by $P_{1}$ and
$W_{1}$.

\end{document}